\documentclass[preprint,12pt,3p]{elsarticle}    

\usepackage{hyperref,amsmath,amsfonts}

\usepackage{amssymb}
\usepackage{amsthm}

\biboptions{sort&compress}

\begin{document}

\begin{frontmatter}

\title{Nonminimal Derivative Coupling Cosmology and the Speed of Gravitational Waves}

\author[a]{Isaac Torres\corref{cor1}}
\ead{its@ufpa.br}
\author[b]{Felipe de Melo Santos}

\cortext[cor1]{Corresponding author}
\address[a]{Faculdade de Física - Universidade Federal do Par\'{a}, Rua Augusto Corrêa n. 01, CEP 66075-110, Bel\'{e}m, PA, Brazil}
\address[b]{Núcleo Cosmo-Ufes - Universidade Federal do Esp\'{i}rito Santo, A. Fernando Ferrari, n. 514, CEP 29075-910, Vit\'{o}ria, ES, Brazil}


\begin{abstract}

The so-called Nonminimal Derivative Coupling (NDC) is an alternative to General Relativity, which produces an asymptotic inflationary mechanism when applied to cosmology. The detection of gravitational waves in the last decade has imposed very stringent constraints over gravitational theories, which gave rise to a massive revision of those theories, in order to investigate the compatibility between them and that observational data. In this paper, we review NDC and address the question if it is compatible with gravitational waves or not. We show that the  very existence of gravitational waves in this theory is restricted to a limited range in phase space and there are no accelerated solutions compatible with the present day data for the speed of such waves. This last result is alleviated by the fact that we did not detect primordial gravitational waves so far. Those conclusions are based on the comparison between the expression for the speed of tensor perturbations and the phase space. Finally, some possible scenarios and solutions are considered.

\end{abstract}


\end{frontmatter}
\section{Introduction}

Scalar-tensor theories became part of the main components of early universe cosmology, since they are present, for instance, in both inflationary and quantum cosmology approaches. Even though there are many covariant theories with a scalar field, the most common one used for this purpose is the so-called minimal coupling, for several reasons \cite{Faraoni_scalar}. This theory consists of a canonical scalar field coupled to gravity, which can be represented by the Lagrangian density below:
\begin{equation}\label{l}
	L[\phi,g_{\mu\nu}] = \sqrt{-g}\left[\frac{R}{16\pi G} -\frac{1}{2}g_{\mu\nu}\nabla^{\mu}\phi \nabla^{\nu}\phi - V(\phi)\right]\;,
\end{equation}
where $R$ is Ricci scalar, $G$ is Newton's constant, $g_{\mu\nu}$ is a metric, $\phi$ is the scalar field and $V(\phi)$ is a scalar potential. 

An important fact about \eqref{l} is that yet a large class of gravitational theories seem quite different from it at first, they are in fact equivalent to it up to a conformal transformation \cite{Faraoni_scalar}. Some interesting examples are $R^2$ inflation, Brans-Dicke theory, and Higgs inflation \cite{Zyla:2020zbs}.

Besides the well known advantadges of canonical scalar field inflation, some of its aspects still need a special attention, such as the initial singularity and the very specific character of the potential $V$, which is considered by some authors as a fine-tuning problem. The singularity problem can be handled with quantum corrections, for instance. For the potential issue, we can aim to avoid it by replacing scalar potential $V(\phi)$ by some other contribution, for example. Thus, if such an alternative theory is not conformally equivalent to \eqref{l}, we can then investigate if it is able to describe an inflationary scenario. If such a structure exists, then it would be an alternative to minimal coupling inflation, without the fine-tuning, for it has no potential. This is one of the main motivations to investigate a scalar-tensor theory substantially different from \eqref{l}.

An important example of such an alternative is the Nonminimal Derivative Coupling, as defined by the covariant Lagrangian below:
\begin{equation}\label{lagr}
	L[\phi,g_{\mu\nu}] =\sqrt{-g}\; \Bigg[\frac{R}{8\pi}-g_{\mu\nu}\nabla^{\mu}\phi \nabla^{\nu}\phi-\kappa G_{\mu\nu}\nabla^{\mu}\phi \nabla^{\nu}\phi \Bigg]\; ,
\end{equation}
where $\kappa>0$ is the non-minimal coupling constant, with dimension of time squared, and $G_{\mu\nu}$ is the Einstein tensor. The main term in \eqref{lagr} is $\kappa G_{\mu\nu}\nabla^{\mu}\phi \nabla^{\nu}\phi$, the coupling between Einstein tensor and the covariant derivatives of the scalar field. This coupling is called nonminimal because it is not conformally equivalent to \eqref{l}, which is a result valid for a broader class of theories, as shown in \cite{AMENDOLA1993175}. This strong distinction between \eqref{l} and \eqref{lagr} is what makes one expect that some new dynamics would follow from \eqref{lagr}. 

Theory \eqref{lagr} and other similar derivative couplings where investigated in various papers, with several different approaches and applications. Most of them (including \eqref{lagr}) can also be seen as particular cases of the bigger Horndeski theory \cite{Horndeski:1974wa,Kobayashi_2019}. The term $G_{\mu\nu}\nabla^{\mu}\phi \nabla^{\nu}\phi$ is also present in the so-called Fab Four theory \cite{PhysRevLett108051101}. Just to mention some of the studies about that kind of couplings, see \cite{PhysRevD81083510,Dent_2013,Babichev2013cya,Gumjudpai2015vio,KAEWKHAO201820,Shahidi2018sas,TORRES2019135003,torres2020quantum}, for instance.

The particular form \eqref{lagr} above was studied in \cite{PhysRevD.80.103505}, where it was shown that it predicts an asymptotic accelerated expansion. This is the main result from \cite{PhysRevD.80.103505}, since it is an indication that an actual inflationary theory could be build up based on that. In this sense, it was shown in \cite{PhysRevD.85.123520} that such a solution matches the duration of inflation if the coupling constant is set to be $\kappa=10^{-74}s^2$.

In this scenario, the revolutionary first detection of gravitational waves have imposed a very strong constraint over their speed $c_{\rm GW}$, which was shown to be really close to that of light \cite{PhysRevLett.119.161101,Goldstein_2017,Abbott_2017}. This means that now all of our theories must deal with this observational data. Thus, this detection was immediately followed by a massive revision of gravitational theories \cite{PhysRevLett.119.251301}. In particular, it was soon realized that NDC and similar nonminimal couplings seem to be incompatible with that constraint \cite{Kobayashi_2019,HEISENBERG20191}. But this does not seem to be an established fact yet, since some authors disagree \cite{Gong2018}.

Hence, it is necessary to analyze if NDC can describe gravitational waves or not. The goal of this letter is to address this question, in the context of cosmology. We show that NDC can describe a gravitational wave with a real-valued speed only for a very restricted range. We show that by studying the dependence of the theoretical value of $c_{\rm GW}$ in terms of the scalar field, which is a consequence of the perturbations in NDC, obtained as a particular case of Horndeski. In summary, we show that outside the interval $-1/(2\sqrt{\pi})<\sqrt{\kappa}\;\dot{\phi}<1/(2\sqrt{\pi})$ we cannot even write down a wave equation for the gravitational waves, according to NDC theory. We also comment the effect of the mentioned constraint from \cite{PhysRevLett.119.161101,Goldstein_2017,Abbott_2017}, which is much more restrictive than the requirement that $c_{\rm GW}$ is a real number.

In Section \ref{revndc}, we first analyze \eqref{lagr} in detail, in two different points of view. We first review that theory, which exhibits four different asymptotic solutions, with a highlight to the inflationary solution proposed in \cite{PhysRevD.80.103505}. Then we introduce a new set of variables that allows us to see the most important asymptotic solutions as points in phase space, from which we can see which ones are sources and attractors.

In Section \ref{gw_section}, we show the precise behavior of $c_{\rm GW}$, which can be expressed in NDC as a function of $\sqrt{\kappa}\dot{\phi}$ only. We first study the speed of gravitational waves in NDC, without taking into account observational data, and then we do it, in great detail. Hence, we can show that only a small range of values for $\sqrt{\kappa}\dot{\phi}$ is capable of producing viable gravitational waves and this range becomes way smaller when present day data is taken into account. Finally, in Section \ref{conc_section} we discuss some possible ways we could make sense of those results.

\section{Nonminimal Derivative Coupling Cosmology}\label{revndc}
Even though there are many theories called derivative couplings, we will refer only to \eqref{lagr} as the Nonminimal Derivative Coupling theory (hereafter called just NDC) through this letter. We now briefly review its basic properties, in the way they were presented in \cite{PhysRevD.80.103505}, but we also present a new asymptotic analysis.

Let us just set some conventions first. We are considering the usual 4-dimensional spacetime, with greek indices running from 0 to 3 and latin indices running from 1 to 3. The background geometry is given by the flat Friedmann-Lema\^itre-Robertson-Walker (FLRW) metric
\begin{equation}\label{flrw}
ds^2 =-N^2dt^2+ a^2\delta_{ij}dx^idx^j\; ,
\end{equation}
where $a(t)$ is the scale factor and $N(t)$ is the lapse function \cite{Calcagni:2017sdq}. We temporarily keep $N$ because it may be useful for future works that involve any quantization process. We are using units such that the speed of light is $c=1$, yet we can write $c$ explicitly sometimes just to emphasize.

From \eqref{flrw}, we find the usual components of the Ricci tensor:
\begin{align}
R_{00} &=3\frac{\dot{a}}{a}\frac{\dot{N}}{N}-3\frac{\ddot{a}}{a}\; ,\\
R_{0i} &=0\; , \\
R_{ij} &=\delta_{ij}\frac{a^2}{N^2}\bigg(2\frac{\dot{a}^2}{a^2}+\frac{\ddot{a}}{a} -\frac{\dot{a}}{a}\frac{\dot{N}}{N}\bigg)\; ,
\end{align}
from which we find the expression for \eqref{lagr} in minisuperspace-like form \cite{bojowald2011quantum}:
\begin{equation}\label{lagr-mini-a}
L=-\frac{3}{4\pi N}a\dot{a}^2+\frac{1}{N}a^3\dot{\phi}^2-\frac{3\kappa}{N^3}a\dot{a}^2\dot{\phi}^2\; ,
\end{equation}
where the dot represents time derivative. The equations of motion have a simpler form if we define $\alpha\equiv\ln a$, and we rewrite \eqref{lagr-mini-a} as
\begin{equation}\label{lagr_mini_alpha}
L=e^{3\alpha}\bigg(-\frac{3\dot{\alpha}^2}{4\pi N} +\frac{\dot{\phi}^2}{N}-\frac{3\kappa \dot{\alpha}^2\dot{\phi}^2}{N^3}\bigg)\; .
\end{equation}

\begin{figure*}[t]
	\centering
	\includegraphics[width=\textwidth]{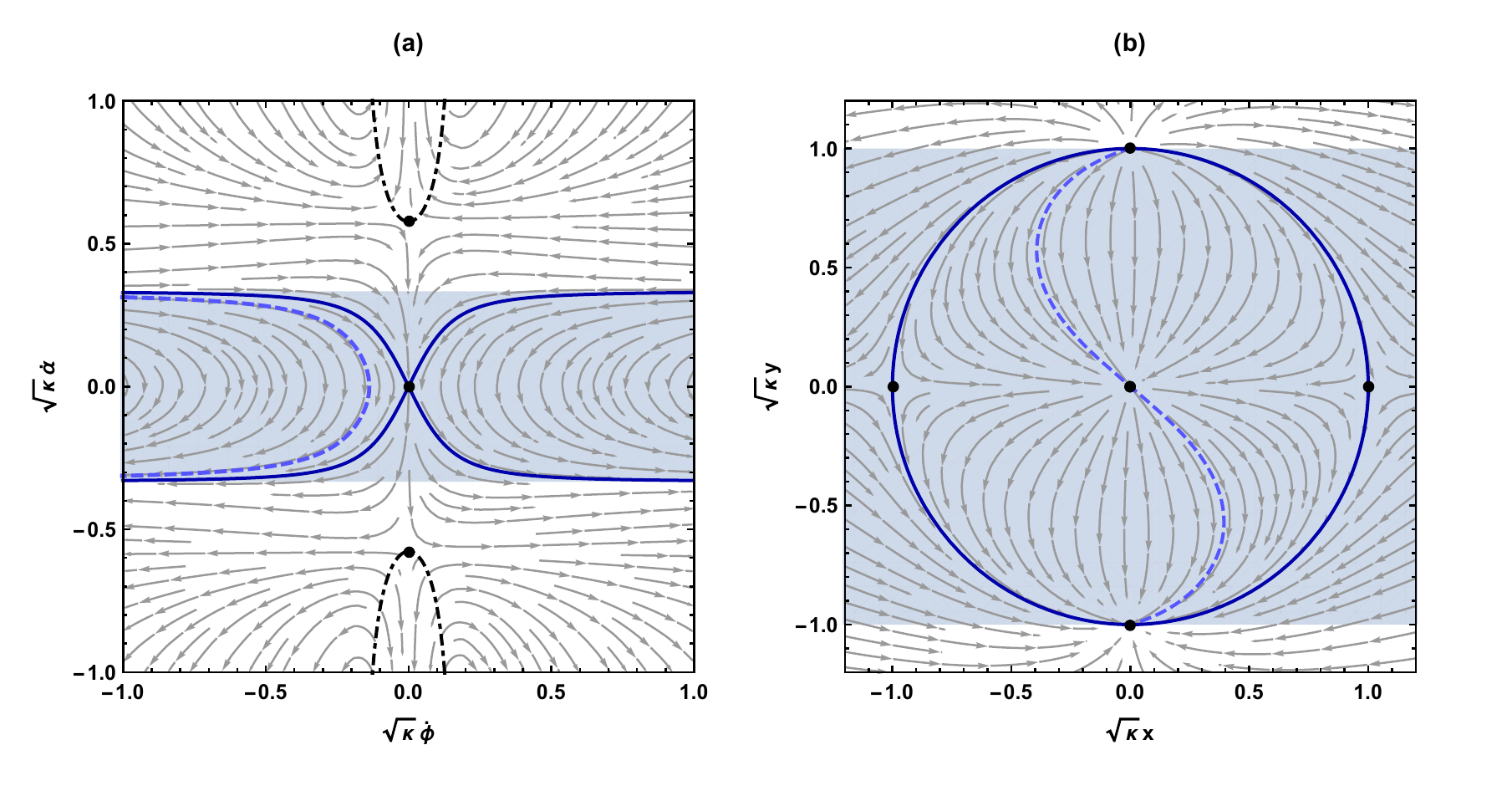}
	\caption{
		The phase portrait (a) represents the dynamical system \eqref{1aeqmovtrue} for $\sqrt{\kappa}\,\dot{\phi}\times\sqrt{\kappa}\,\dot{\alpha}$. The direction of the arrows indicates time evolution; the three black dots represent the critical points \eqref{crpointsaf}, all of which are unstable; the thick curves in blue represent the constraint \eqref{constr}; the dashed blue curve is just an example of solution, for which the initial conditions are $\dot{\alpha}(0)=0.1/\sqrt{\kappa}$, $\dot{\phi}(0)=-1/\sqrt{12\pi\kappa}$. The shaded region represents the allowed range of values  $|\dot{\alpha}|<1/\sqrt{9\kappa}$, a limitation imposed by \eqref{constr}; the dash-dotted lines represent the singularities, for which the denominator of \eqref{1aeqmovtrue} vanishes. As for phase portrait (b), it represents the dynamical system \eqref{dsxy}, where the variables are now $x$ and $y$, defined in \eqref{defxy}. The shaded region is the range of values $|y|<1$, which is equivalent to $|\dot{\alpha}|<1/\sqrt{9\kappa}$. The black dots represent the critical points relevant for the present discussion, which are $(0,0)$, $(\pm1,0)$, and $(0,\pm1)$ (the other critical points lie outside the constraint circle); the blue thick line represents the constraint \eqref{constrxy}; the dashed line is an example of solution, with initial conditions $x(0)=-0.3$, $y(0)=0.3$, which is the same example shown in (a).
	}
	\label{fig1}
\end{figure*}

Now we can study the equations of motion which describe the cosmological evolution generated by NDC, on the minisuperspace form \eqref{lagr_mini_alpha}. The variation with respect to $N$ gives the constraint below, setting $N=1$ after deriving the equation:
\begin{equation}\label{constr}
	\frac{3}{4\pi}\dot{\alpha}^2-\dot{\phi}^2+9\kappa\dot{\alpha}^2\dot{\phi}^2=0\; .
\end{equation}
The variation with respect to $\alpha$ and $\phi$, respectively, gives the equations of motion below, for $N=1$:
\begin{subequations}\label{1aeqmov-class}
\begin{align}
	2\ddot{\alpha}+3\dot{\alpha}^2+4\pi\dot{\phi}^2[ 1+\kappa(2\ddot{\alpha}+3\dot{\alpha}^2+4\dot{\alpha}\ddot{\phi}\dot{\phi}^{-1})] &=0\;,		\label{1aeqmov-class-alpha}\\
	\ddot{\phi}+3\dot{\alpha}\dot{\phi}-3\kappa(\dot{\alpha}^2\ddot{\phi}+2\dot{\alpha}\ddot{\alpha}\dot{\phi}+3\dot{\alpha}^3\dot{\phi}) &=0\;.		\label{1aeqmov-class-phi}
\end{align}
\end{subequations}

Equations \eqref{constr}, \eqref{1aeqmov-class-alpha}, and \eqref{1aeqmov-class-phi} were derived in \cite{PhysRevD.80.103505}. Solving for $\ddot{\alpha}$ and $\ddot{\phi}$, the second-order system \eqref{1aeqmov-class} can be rewritten as
\begin{subequations}\label{1aeqmovtrue}
	\begin{align}
	\ddot{\alpha}&=	-\frac{3\dot{\alpha}^2-9\kappa\dot{\alpha}^4+4\pi\dot{\phi}^2-48\pi\kappa \dot{\alpha}^2\dot{\phi}^2+108\pi\kappa^2\dot{\alpha}^4\dot{\phi}^2}{2( 1-3\kappa\dot{\alpha}^2 +4\pi\kappa\dot{\phi}^2+36\pi\kappa^2\dot{\alpha}^2\dot{\phi}^2)}\; ,\label{1aeqmovtrue_alpha}\\
\ddot{\phi}&=-\frac{3\dot{\alpha}\dot{\phi}(1+8\pi\kappa\dot{\phi}^2)}{1-3\kappa\dot{\alpha}^2 +4\pi\kappa\dot{\phi}^2+36\pi\kappa^2\dot{\alpha}^2\dot{\phi}^2}\; .\label{1aeqmovtrue_phi}
	\end{align}
\end{subequations}

Note that this system is not the same originally presented in \cite{PhysRevD.80.103505}, but they are equivalent. Now, we start analysing NDC theory based on \eqref{1aeqmovtrue}. The structure of system \eqref{1aeqmovtrue} shows that it can be seen as an autonomous dynamical system for which $\dot{\alpha}$ and $\dot{\phi}$ are the independent variables. The phase portrait of that system is shown in Fig. \ref{fig1}. Let us first discuss critical points and asymptotic solutions, because this leads to the inflationary mechanism claimed in \cite{PhysRevD.80.103505}.

There are three relevant critical points $(\dot{\phi}_{\rm C},\dot{\alpha}_{\rm C})$ of \eqref{1aeqmovtrue}:
\begin{equation}\label{crpointsaf}
	(0,0)\;,\quad (0,1/\sqrt{3\kappa})\;,\quad\mbox{and}\quad (0,-1/\sqrt{3\kappa})\;.
\end{equation}
The point $(0,0)$ is a trivial solution. The point $(0,1/\sqrt{3\kappa})$ represents an exponential acceleration, because $\dot{\alpha}$ is the same as the Hubble factor. And $(0,-1/\sqrt{3\kappa})$ represents an exponential deceleration.

After some algebra, we can identify four asymptotic solutions for the scale factor:
\begin{enumerate}
	\item[(S1)] for $|\dot{\phi}|\longrightarrow\infty$ and $\dot{\alpha}>0$, $\; a\sim e^{t/\sqrt{9\kappa}}$;
	\item[(S2)] for $|\dot{\phi}|\longrightarrow\infty$ and $\dot{\alpha}<0$, $\; a\sim e^{-t/\sqrt{9\kappa}}$;
	\item[(S3)] for $\dot{\phi}\longrightarrow0$, $\; a\sim t^{2/3}$;
	\item[(S4)] for $\kappa\longrightarrow0$, $\; a\sim t^{1/3}$;
\end{enumerate}

Let us now show how those solutions can be obtained by taking some limits on \eqref{1aeqmovtrue}. Solutions (S1) and (S2) are obtained by rewriting \eqref{constr} as 
\begin{equation}
	\frac{3}{4\pi}\frac{\dot{\alpha}^2}{\dot{\phi}^2}-1+9\kappa\dot{\alpha}^2=0
\end{equation}
and then taking $|\dot{\phi}|\longrightarrow\infty$; (S3) comes from the simplification of \eqref{1aeqmovtrue_alpha} when $\dot{\phi}\longrightarrow0$; finally, (S4) is just the canonical scalar field solution when there is no potential. The phase portrait on the left handed side of Fig. \ref{fig1} helps us interpreting those solutions.

The solution (S1) describes the asymptotic behavior in the early times for a solution like the one represented by the blue dashed line, and also both sides of the upper half of the constraint line in Figure \ref{fig1} (a). This is the inflationary solution presented in \cite{PhysRevD.80.103505}, where the coupling constant is set to be $\kappa=10^{-74}s^{2}$, so that it fits in the duration of inflation, as explained in \cite{PhysRevD.85.123520}. Note that, in any case, all solutions like (S1) lie inside the allowed $\dot{\alpha}$ range, which is $-1/\sqrt{9\kappa}<\dot{\alpha}<1/\sqrt{9\kappa}$, according to \eqref{constr}. Hence, if we follow the path of any solution like (S1), it comes from $t\longrightarrow-\infty$ by an asymptotic de Sitter solution, without breaking the condition $-1/\sqrt{9\kappa}<\dot{\alpha}<1/\sqrt{9\kappa}$. Thus, (S1) represents the inflationary solution constructed in \cite{PhysRevD.80.103505}.

As for the other solutions: from (S2), a contracting universe is possible according to NDC; from (S3), there is a matter-dominated era when $\dot{\phi}\approx0$; finally, we see from (S4) that the universe is dominated by stiff matter when the nonminimal coupling term is negligible.


Some of those asymptotic solutions become points if we define convenient variables. Let us take
\begin{equation}\label{defxy}
x\equiv\sqrt{\frac{3}{4\pi}}\frac{\dot{\alpha}}{\dot{\phi}}\;,\qquad y\equiv\sqrt{9\kappa}\,\dot{\alpha}\;.
\end{equation} 
Now, as we shall see in more detail below, the asymptotic solutions (S1) and (S2) become points in phase space. Thus, we gain new information, since now we can easily see which asymptotic solution is an attractor and which one is a source.

For $x$ and $y$, constraint \eqref{constr} becomes the unit circle
\begin{equation}\label{constrxy}
x^2+y^2=1
\end{equation}
and the equations of motion, after some algebra, become the following dynamical system:
\begin{subequations}\label{dsxy}
	\begin{align}
	\dot{x}&= -\frac{xy\, (\, y^4-x^2y^2-8y^2-3x^2+3\, )}{2\sqrt{\kappa}\,(\, y^4-x^2y^2+y^2+3x^2\, )}	\;,\\
	\dot{y}&= -\frac{y^6-x^2y^4-4y^4+3x^2y^2+3y^2}{2\sqrt{\kappa}\,(\, y^4-x^2y^2+y^2+3x^2\, )}	\;.
	\end{align}
\end{subequations}
The phase portrait of \eqref{dsxy} is shown in Fig. \ref{fig1} (b). Let us now study its main critical points and solutions.

The critical points $(x_{\rm C},y_{\rm C})$ of \eqref{dsxy} are
\begin{equation}\label{crpointsxy}
(0,0)\;,\quad (\pm1,0)\;,\quad (0,\pm1)\;,\quad\mbox{and}\quad (0,\pm\sqrt{3})\;.
\end{equation}
From Figure \ref{fig1} (b), $(0,1)$ is an unstable node and $(0,-1)$ is an attractor. Thus, from \eqref{defxy}, we can see the actual meaning of those points: $(0,1)$ corresponds to (S1), the de Sitter solution $\; a\sim e^{t/\sqrt{9\kappa}}$; hence it is a source of dynamical solutions in the phase space.

We find solution (S2), which is $\; a\sim e^{-t/\sqrt{9\kappa}}$, at the point $(0,-1)$ in the $xy$ plane. As we can see in Figure \ref{fig1} (b), this is the final attractor. Note that both the asymptotic limits $\dot{\phi}\longrightarrow\pm\infty$ are condensate in $x=0$ (for any $\dot{\alpha}\neq0$). 

In particular, then, we can take the dashed curve in Figure \ref{fig1} (b), which corresponds to a dynamics that comes from $|\dot{\phi}|\longrightarrow+\infty$, passes through $\dot{\phi}=0$ and then evolves back to $|\dot{\phi}|\longrightarrow+\infty$, and this last behavior is actually an attractor, meaning that all relevant solutions necessarily evolve to $|\dot{\phi}|\longrightarrow+\infty$. In the next Section, we will discuss the precise relation between this particular behavior and gravitational waves. 

As a final comment in this Section, we should mention that, since the origin in the $xy$ phase space is a critical point, in principle a solution that goes to $(0,0)$ never actually comes from this state. But this is not true. Take, for instance, the blue dashed curve in Figure \ref{fig1} (a). For the original variables $\sqrt{\kappa}\dot{\phi}$ and $\sqrt{\kappa}\dot{\alpha}$, we virtually see all its behavior: it asymptotically comes from (S1) and then, after it approaches $\dot{\alpha}=0$, it starts to evolve to the asymptotic solution (S2). And there is no singularity in between. Thus, when we write the same solution, the blue dashed curve, in terms of $x$ and $y$, the point $(0,0)$ in $xy$ plane is not an actual physical critical point, it is just a consequence of the definition of the variables $x$ and $y$ (see \eqref{defxy}). In other words, it does make sense to say that the solutions inside the unitary disk in Fig. \ref{fig1} (b) come from the source point $(0,1)$, which represents (S1), and then they evolve to the attractor $(0,-1)$, which represents (S2), as we claimed above.

\section{Gravitational Waves Constraint over NDC}\label{gw_section}
As we know, the recent detection of gravitational waves has imposed a very tight constraint over gravitational theories \cite{PhysRevLett.119.161101,Goldstein_2017,Abbott_2017}:
\begin{equation}\label{cap1-eq-vinc-gw}
-3\times10^{-15}<c_{\text{GW}}/c-1\leq7\times10^{-16}\;.
\end{equation}
This was a revolutionary measure that showed how close the observed value of $c_{\text{GW}}$ is to one, which is the prediction of general relativity. In this Section, we will investigate ways to decide if NDC can describe gravitational waves. After that, we will focus on the relation between \eqref{cap1-eq-vinc-gw} and NDC.

Notice that NDC is a particular case of Horndeski theory \cite{Horndeski:1974wa}, which is represented by the action
\begin{equation}\label{horndact}
S_H[\phi,g_{\mu\nu}]=\int d^4x\sqrt{-g}(L_2+L_3+L_4+L_5)\;,
\end{equation}
where
\begin{align}\label{hrndskip}
L_2 & =K(\phi,X)\;,  \\
L_3 & =-G_3(\phi,X)\Box\phi\;,  \\
L_4 & = G_4(\phi,X)R+G_{4,X}(\phi,X)[(\Box\phi)^2  \nonumber\\ &-\nabla^{\mu}\nabla^{\nu}\phi\nabla_{\mu}\nabla_{\nu}\phi]\;,\\
L_5 & = G_5(\phi,X)G^{\mu\nu}\nabla_{\mu}\nabla_{\nu}\phi-\textstyle\frac{1}{6}G_{5,X}(\phi,X) [(\Box\phi)^3\nonumber\\
& -3\Box\phi\nabla^{\mu}\nabla^{\nu}\phi\nabla_{\mu}\nabla_{\nu}\phi   \nonumber\\ &+2\nabla_{\mu}\nabla_{\nu}\phi\nabla_{\lambda}\nabla^{\mu}\phi\nabla^{\nu}\nabla^{\lambda}\phi]\;.
\end{align}
The functions $K(\phi,X)$ and $G_i(\phi,X)$ are generic; $X\equiv-\frac{1}{2}\nabla^{\mu}\phi\nabla_{\mu}\phi$ is the kinetic term; the notation $G_{i,X}$ represents the derivative of $G_{i}$ with respect to $X$. Thus, we see that NDC is the particular case of Horndeski for which $K(\phi,X)=2X$, $G_3=0$, $G_4=1/8\pi$, and $G_5=G_5(\phi)$ is such that $dG_5/d\phi=\kappa$. This last equality is obtained by integrating by parts the $G_5$ term of Horndeski when $G_5=G_5(\phi)$ and comparing it with \eqref{lagr}, discarding surfaces terms.

\begin{figure}[t]
	\centering     
	\includegraphics[width=7cm]{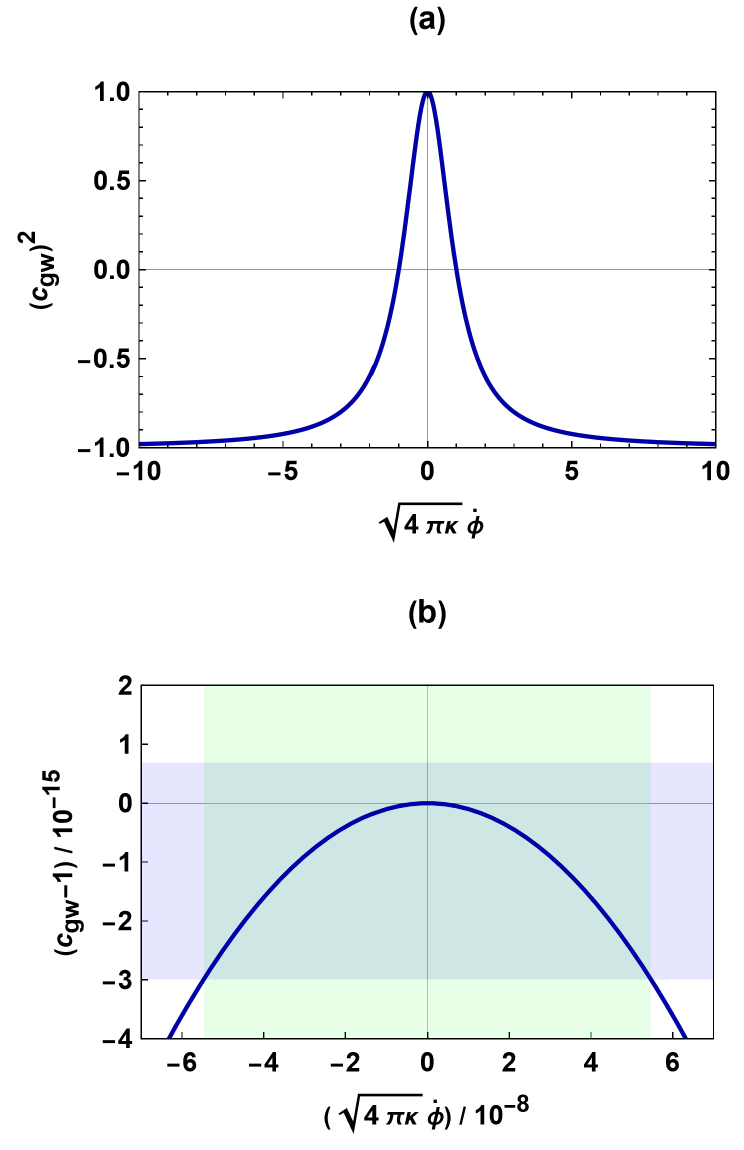}\caption{Speed of gravitational waves for NDC according to Eq. \eqref{c2gwsush}. Plot (a) represents a wide range of values of $c_{GW}^{2}$ (where $1=c$) as a function of $\sqrt{4\pi\kappa}\;\dot{\phi}$. In plot (b), we zoom in plot (a) to show the relation between \eqref{c2gwsush} and \eqref{cap1-eq-vinc-gw}: the blue shaded region represents \eqref{constraint_ligo}, and the green shaded one represents \eqref{constrsushgw}.}
	\label{fig2}
\end{figure}

For reviews of Horndeski theory, see \cite{Kobayashi_2019,HEISENBERG20191}. For the present purposes, we must remember that Horndeski theory is the most general covariant scalar-tensor gravitational theory in four dimensions with second-order equations of motion. This means that in four dimensions, including only one extra degree of freedom, which is the scalar field $\phi$, under very reasonable assumptions we are lead unavoidably to Horndeski theory. The second-order equations of motion guarantee that Ostrogradsky's instability is avoided. Hence, in particular, those are features of NDC as well.

The perturbative analysis of \eqref{horndact} was first developed in \cite{Felice_2012} and then in \cite{Bellini_2014} a deeper comprehension of those perturbations and their relation with structure formation (with lots of particular cases and examples) was presented. According to \cite{Bellini_2014}, the squared speed of gravitational waves in Horndeski gravity is given by:
\begin{equation}\label{c2gw}
c_{\text{GW}}^{2}=\frac{G_4-X(\ddot{\phi}G_{5,X}+G_{5,\phi})}{G_4-2XG_{4,X}-X(H\dot{\phi}G_{5,X}-G_{5,\phi})}\;,
\end{equation}
which, for the present case, reduces to a much simpler form:
\begin{equation}\label{c2gwsush}
	c_{\text{GW}}^{2}=\frac{1-4\pi\kappa\dot{\phi}^2}{1+4\pi\kappa\dot{\phi}^2}\;.
\end{equation}

In Figure \ref{fig2} (a), we can see $c_{\text{GW}}^{2}$ as a function of $\sqrt{4\pi\kappa}\;\dot{\phi}$. Since $\kappa>0$, the last expression immediately implies that
\begin{enumerate}
	\item[(i)] if $-1/(2\sqrt{\pi})<\sqrt{\kappa}\;\dot{\phi}<1/(2\sqrt{\pi})$, then we can define $c_{\text{GW}}$ as a positive real quantity, which is a function of $\sqrt{\kappa}\;\dot{\phi}$;
	\item[(ii)] if $\sqrt{\kappa}\;\dot{\phi}=\pm 1/(2\sqrt{\pi})$, then $c_{\text{GW}}=0$;
	\item[(iii)] if $\sqrt{\kappa}\;\dot{\phi}<-1/(2\sqrt{\pi})$ or $\sqrt{\kappa}\;\dot{\phi}>1/(2\sqrt{\pi})$, then $c_{\text{GW}}$ is purely imaginary.
\end{enumerate}




Because of the wave equation, we can interpret the above classification in the following way: (i) represents a standard wave equation with speed $c_{\text{GW}}>0$; (ii) would lead to a degenerate equation and (iii) would lead to a Laplace equation, which is elliptic, instead of hyperbolic, like the wave equation. In other words, the only set of values capable of describing the tensor perturbations as physical waves is (i). 

\begin{figure}[h]
	\centering
	\includegraphics[width=7cm]{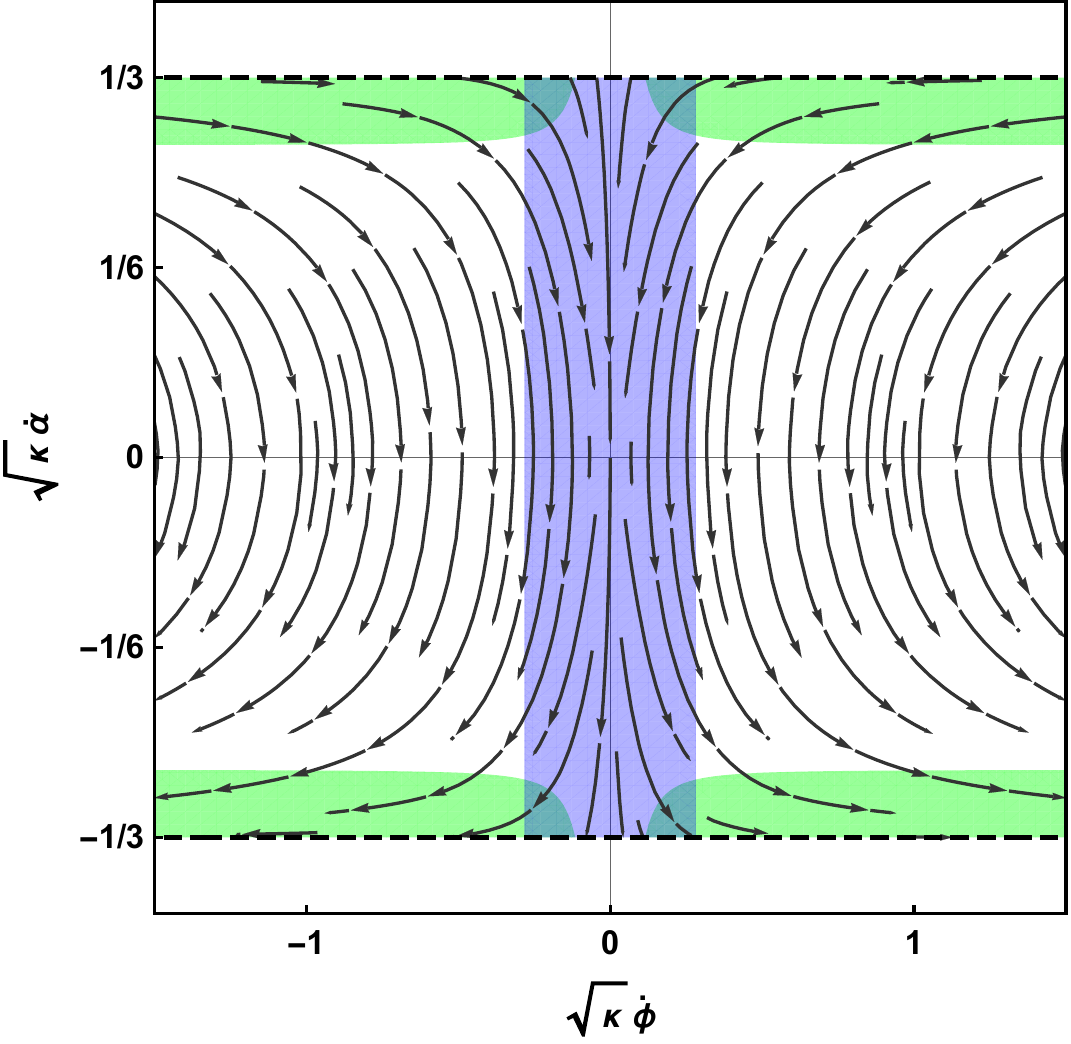}
	\caption{Comparison between two regions in phase space: the green shaded region represents points where there is acceleration and the blue shaded region (excluding its boundary $\sqrt{\kappa}\;\dot{\phi}=\pm 1/(2\sqrt{\pi})$) represents the states for which there are gravitational waves propagating with a positive speed obtained from \eqref{c2gwsush}.
	}
	\label{fig3}
\end{figure} 

Now we can return to phase space, in Figure \ref{fig3}: the vertical blue range represents the interval $-1/(2\sqrt{\pi})<\sqrt{\kappa}\;\dot{\phi}<1/(2\sqrt{\pi})$. Thus, inside that region, there are gravitational waves whereas outside NDC cannot describe such waves. We can understand better this behavior by studying the time evolution of some cosmological solution: any given trajectory always comes from outside the blue range, where there are no gravitational waves, and then it evolves to near $\sqrt{\kappa}\dot{\alpha}=0$, where it may or may not pass through the blue region, depending on the initial conditions.

If the solution does not intersect the blue shaded region, this means an eternal universe without gravitational waves at all. If it passes through the blue region, this means there is a time interval when gravitational waves can propagate, with a variable speed (determined by \eqref{c2gwsush}) with maximum value 1 (in c units) and at some point the solution leaves the blue shaded region. At this point, gravitational waves suddenly stop existing again.

Observe that there are four intersections in Figure \ref{fig3}, which represent the points for which there is expansion with gravitational waves propagating with a well defined speed. If some given solution will or will not pass through theses regions depends only on the initial conditions, as shown in Figure \ref{fig3}.

Notice that those conclusions apply to any $\kappa>0$, so the same arguments are valid, in particular, for $\kappa=10^{-74}s^2$, the value used in \cite{PhysRevD.85.123520}. All those limitations follow immediately from Eq. \eqref{c2gwsush}, the very definition of the squared speed of gravitational waves in NDC.


If we take into account the detection of  gravitational waves, which leads to \ref{cap1-eq-vinc-gw}, there seems to exist an even tighter constraint:
\begin{equation}\label{constraint_ligo}
	1-3\times10^{-15}<\Bigg(\frac{1-4\pi\kappa\dot{\phi}^2}{1+4\pi\kappa\dot{\phi}^2}\Bigg)^{1/2}\leq1+7\times10^{-16}\;.
\end{equation}
The inequality on the right-hand side is automatically satisfied since the maximum value of \eqref{c2gwsush} is 1, for any $\dot{\phi}$ and any $\kappa>0$. Solving the inequality on the left-hand side for $\dot{\phi}$, we find:
\begin{equation}\label{constrsushgw}
	\sqrt{4\pi\kappa}\;|\dot{\phi}|\lesssim\sqrt{30}\times 10^{-8}\;.
\end{equation}
In Figure \ref{fig2} (b), we can see this range (green region) and the correspondent values of the deviation $c_{\textrm GW}-1$ (blue region).  The intersection of the two shaded regions represents the range compatible with both \eqref{cap1-eq-vinc-gw} and \eqref{constrsushgw}. We will discuss first the relation of that constraint with the solutions and then we will analyze its applicability.

From the previous Section we see that any solution of the system \eqref{1aeqmovtrue} goes to or comes from $|\dot{\phi}|\longrightarrow\infty$. Therefore, they all break \eqref{constrsushgw}, at some point. In other words, there cannot be an asymptotic inflationary solution of NDC in accordance with \eqref{constrsushgw}. In particular, the inflationary solution for which $\kappa=10^{-74}s^2$, doesn't satisfy \eqref{constrsushgw} either. It is true that when $\kappa$ gets smaller, the constrained range for $|\dot{\phi}|$ becomes wider, in view of \eqref{constrsushgw}. However, no matter how big the range of $|\dot{\phi}|$ is, there is an unavoidable incompatibility between the limitation imposed by this range and the fact that any inflationary solution is taken asymptotically when $|\dot{\phi}|$ goes to infinity.

Nevertheless, there may be other accelerated solutions, in principle. If we consider the phase space $\sqrt{\kappa}\dot{\phi}\times\sqrt{\kappa}\dot{\alpha}$, we can map the points where there is acceleration ($\ddot{a}>0$). Since $\ddot{a}/a=\dot{\alpha}^2+\ddot{\alpha}$, there will be acceleration when the quantity $\dot{\alpha}^2+\ddot{\alpha}$ is positive. As one can check, we can express the quantity $\kappa\ddot{a}/a$ as a function of $\sqrt{\kappa}\dot{\phi}$ and $\sqrt{\kappa}\dot{\alpha}$ only, with no explicit dependence on $\kappa$, and $\kappa\ddot{a}/a>0$ if, and only if, $\ddot{a}>0$.

In other words, the points in phase space where there is acceleration are precisely the ones for which the quantity $\kappa\ddot{a}/a$ is positive and we can express this for any $\kappa>0$. In Figure \ref{fig3}, we show those points represented in the green shaded region. Now we can compare this set of points with the ones for which constraint \eqref{constrsushgw} is valid. But, since \eqref{constrsushgw} determines only a very thin vertical range around $\sqrt{\kappa}\dot{\phi}=0$ (which is $8$ orders of magnitude thinner than the blue shaded range), there is actually no intersection, which means there are no accelerated solution compatible with  \eqref{constrsushgw}.

We have thus shown that in Nonminimal Derivative Coupling the existence of a wave equation for tensor perturbations is restricted to a limited range in phase space. We also have shown that there are no accelerated solutions compatible with the measured value of the speed of gravitational waves known today.

Since no primordial gravitational waves were detected so far, in principle, one can argue that the experiments are not enough to show that the speed measured today applies to a remote past, where both asymptotic inflation and primordial gravitational waves were supposed to coexist. Let us consider some possibilities on the Conclusion.

\section{Conclusion}\label{conc_section}
There are basically three possible scenarios for the future, in order to give a final decision to the problem addressed here: (A) if in the future those primordial waves are detected and their speed matches the today known value, we will conclude that NDC is ruled out, by the discussion at the end of the previous Section; (B) if we detect those waves and their speed is significantly different from 1, then NDC is also ruled out, simply because there are no primordial gravitational waves in NDC. This is so because the further we go back in time in NDC the further we go away from the range were gravitational waves exist; (C) if we somehow find out through observations that there are no such waves, then NDC would still be a viable cosmological theory.

So the viability of that derivative coupling is tied to a particular kind of description of gravitational waves that is still to be clarified by future observations, namely the possibility that gravitational waves exist only in a some specific range of values of $\sqrt{\kappa}\dot{\phi}$.

It is also possible that to understand better the description of NDC for primordial gravitational waves, one should take into account quantum effects. We aim to address such possibility in a future work.

\section*{Acknowledgments}
We thank to Felipe Tovar Falciano, Ingrid Ferreira da Costa, J\'ulio C\'esar Fabris, Nelson Pinto-Neto, Paola Carolina Moreira Delgado, and Tays Miranda de Andrade for very important discussions about this paper. We also thank the anonymous Reviewer who raised an important question whose answer complemented our result. This study was financed in part by the \emph{Coordena\c{c}\~ao de Aperfei\c{c}oamento de Pessoal de N\'ivel Superior} - Brazil (CAPES) - Finance Code 001, and also by CNPq and FAPES, from Brazil.

\bibliography{mybibfile}
\end{document}